\documentclass{article}

\usepackage{arxiv}

\usepackage[utf8]{inputenc} 
\usepackage[T1]{fontenc}    
\usepackage{hyperref}       
\usepackage{url}            
\usepackage{booktabs}       
\usepackage{amsfonts}       
\usepackage{nicefrac}       
\usepackage{microtype}      
\usepackage{caption}
\usepackage{graphicx}
\usepackage{multirow}
\usepackage{url}

\begin{document}
\begin{center}
\title{Why Monday never wins \\ An Example of the Secretary Problem}
\end{center}

\author{
  Peter Blum\thanks{peter.blum@desy.de}\\ 
  Institute of Experimental Physics\\
  Universit\"at Hamburg
   \And
 Marc Wenskat\\Institute of Experimental Physics
\\Universit\"at Hamburg
}

\maketitle

\begin{abstract}

We analyzed the winning statistics of the German TV show 'Das Perfekte Dinner', a competition where the contestants rate each other. We did a comparison of the original and the celebrity version of the show and also implemented a rescaling of the points to reduce the influences of subjective grading. We found that there is a strong dependency of the winning rate on the position of performing, not only for the first, but for all candidates. Furthermore, we concluded that there is a positive bias in giving points, after the contestants have already cooked. The comparison with a simpler optimizing strategy, the so called secretary problem, showed a lot of similarities.

\end{abstract}


\section{Introduction}

There are a lot of different competitions in the world as people want to find out who is the best. And there are as many rating systems as competitions. One of them, which is commonly used, is that judges award points to the contestants in a certain range, typically from 1 to 10,immediately after their performance. Whereas for example in sports these ratings depend on difficulty or accuracy, in more abstract competitions like poetry slams or cooking competitions they solely depend on personal taste of the judges.\\
The question occurs how fair such ratings are. One commonly discussed phenomenon is the 'order effect', which states that the winning probability depends on the order of performing. As there are no sufficiently large data sets on voting at science slams, we looked at a different  show with a similar voting behavior : 'Das perfekte Dinner'.\\
This TV show has the difference that the competitors rate each other, but otherwise the voting system is very similar. There already is some research on the voting in 'Das Perfekte Dinner', but none with a data set as large as ours.\\

We will try to compare the 'order effect' with the secretary problem, a very simple rating problem which only uses one judge who can accept or reject a candidate and develops the strategy to finding the best candidate. The data shows significant similarity with this strategy, although the contestants supposedly not know it.\\
Furthermore, we found contradictions to the existing papers, especially that the 'order effect' is only regarding the first contestant.  We agree that the first contestant has the least chances but there is also a dependency for the later contestants. We find that it is the best to perform at late as possible to maximize the chances of winning.\\

Another dependency introduced by \cite{schueller} is a negative 'cooking bias', stating that a contestant who has already performed will give less points on average. We tried to prove this fact not with the method of multi-linear regression but with averaging before and after cooking.\\

In this paper, we will first describe the show 'Das Perfekte Dinner' and then introduce the secretary problem. The method of data aquisition will be stated as well as the procedure of the data. Section 6 presents the results, which will be discussed afterwards. 

\section{The Show "Das Perfekte Dinner"}
The show is based on a British format called "Come Dine with Me". The German version is being broadcasted since 2006 regularly (after a single pilot show in 2005) and an overall of 3351 episodes aired\textsuperscript{\cite{wiki_DPD}}. 
\newline
The concept of the show is that every week a group of five strangers, each week with new candidates, act as hosts for each other. Each host will cook once, on a weekday, a three course menu for the others. To win the week one has to be the "perfect host" - or closest to one. To determine this, after the menu of the candidate, the other four candidates vote on a scale of 0 to 10, where 10 is a perfect dinner, and the sum of the individual votings is the score of the host. The scores are kept secret until the end of the week. With this system, multiple winner are possible. The winner(s) of the week will win 3000 Euros, the other candidates will win nothing. In case of a short week (because of e.g. a holiday), only four candidates will compete.
\newline
In addition, several modifications of the show have aired, such as one candidate having a sleep over at the host's or a version in which a professional cook in disguise competes with the amateurs.
Another interesting version, where the candidates are celebrities, is also broadcasted since 2006. 

\section{The Secretary Problem}
The problem was first stated in a column of the \textit{Scientific American} from 1960 by Martin Gardner and has been reformulated since then several times\textsuperscript{\cite{secretary1}}.
The common rules are\textsuperscript{\cite{secretary2}}: 
\begin{itemize}
    \item There is a single position to fill
    \item There are N candidates, where N is known
    \item An unambiguous ranking of the candidates (best to worst) is possible
    \item The candidates are interviewed sequentially
    \item After each interview, the candidates have to be rated immediately, i.e. accepted or rejected
    \item The decision stands - the candidates leave the application process and can not return
\end{itemize}
It has been shown that a stopping rule is the optimal strategy for this problem. For large N, the probability to choose the best candidate converges against an optimal value of 1/e $\approx$ 36.7\% if the first n/e candidates are rejected and serve as a reference group. The other candidates are chosen, if they are better than the reference group. For small values of N, the optimal threshold for rejection and the resulting probability to find the best candidates can be found using suitable algorithms\textsuperscript{\cite{Beckmann}}. The resulting threshold and probability for 5 contestants is 3 and 43.3\% and for 4 contestants 2 and 45.8\%.
\newline
There is a variant of this problem called "cardinal payoff variant", in which simply someone from the upper tier of candidates should be chosen compared to the lower tier\textsuperscript{\cite{Bearden}}. The threshold is then calculated to skip the first $\sqrt{n} - 1$ candidates, which effectively results in an earlier acceptance of candidates for most values of n. Already in our case of n\,=\,5 or 4 there is a difference as only the first candidate should be rejected compared to 3 or 2 rejections for the original formulation.
\newline
This decision problem is similar to the voting system of the TV show described above, except there are three modifications
\begin{enumerate}
    \item There is no clear rejection or acceptance of a candidate. Due to the rating system, the immediate decision is not final in the context, as for example the first vote can only be interpreted as an acceptance or rejection in comparison to the later votes.
    \item The judges are the candidates themselves
    \item The ranking can be ambiguous, e.g. multiple "best cooks" and a subjective grading of a judge
\end{enumerate}

\section{Method}
\label{sec:method}
A simple descriptive statistical approach was chosen similar to \textsuperscript{\cite{haigner}}. In addition we implemented a rescaling of the votes to take into account two effects playing a role leading to a subjective grading:
\begin{enumerate}
    \item The same dinner, which is perceived by two different contestants the same way as "good", might still get two different grades, e.g. 6 and 7, from those two contestants. So different contestants use different means and scalings in their ratings.
    \item Different backgrounds can create a skewness in the grading system. A more critical person might give on average less points because he is used to a higher quality and his ceiling is higher, hence his distribution would have a positive skewness. Another person might be used to different level of quality and would expect more less-then average food and hence pronounces or stretches the lower grades and have a negative skewness. 
\end{enumerate}
To minimize the influence of these effects, we normalized the average points given by each cook that week to a set-point of 5.
$$ p_{i,new} = \frac{p_{i,orig}}{\bar{p}} \cdot 5 $$
In the formula $p_{i,new}$ is the new number of points awarded, $p_{i,orig}$ is the original number of points awarded, and $\bar{p}$ is the average points awarded by the cook originally.\\
Using this algorithm, we minimized the influence of the different averages between the different cooks, hence addressing the first mentioned issue. The scaling and skewness differences can not be removed as at most 5 ratings of an individual contestant are not enough to properly fit any distribution to the ratings.\\
The data of the celebrity version only had the sum of each contestant, hence this re-scaling was not possible on that data set.  

\section{Data}
The data used in this study is a sample of 2268 regular episodes broadcasted from 2006 to 2019. An additional data set of 540 episodes of the celebrity version is used. The data was taken from the German Wikipedia article\textsuperscript{\cite{wiki_DPD}} for the celebrity version and from an online-forum\textsuperscript{\cite{blog}} for the regular version. The values shown will be given as mean $\pm$ standard error of the mean.
We split the data set of the regular show since 9 \% of the weeks have been "short weeks", meaning only four shows took place and four contestants competed.
From the data available for the regular version, we excluded 75 weeks where special rules have been applied, such as extra-ratings for the decoration or having one professional cook in disguise among the contestants. Furthermore, we also excluded champion-episodes, where winners of other weeks cooked against each other and two weeks have been excluded due to insufficient data. An overview of the data set is given in \autoref{tab:data}\\
\begin{table}[!hbt]
   \centering
   \caption{Overview of data used.}
   \vspace{0.5cm}
\begin{tabular}{lcccccc}
      \toprule
    	Show	& Label  & cooks/week & \#weeks & \#winners & \#winners & avg. points  \\
			    &        &            &        &          & rescaled &              \\
	   \cline{1-7}
	   Das Perfekte Dinner & PD5 & 5 & 420 & 492 & 421 & $7.57\,(\pm\,0.06)$\\	   
       Das Perfekte Dinner & PD4 & 4 & 42 & 49 & 44 & $7.55\,(\pm\,0.24)$\\	   	                          
       Das Perfekte Promi-Dinner & PDP & 4 & 135 & 174 & x & $8.05\,(\pm\,0.13)$\\	   	                          
		\bottomrule
   \end{tabular}
   \label{tab:data}
\end{table}

Compared to our data set of 597 weeks from 2006 to 2019, the data set from \textsuperscript{\cite{haigner}} included only 186 weeks from March 2006 to February 2010 and the data set from \textsuperscript{\cite{schueller}} included 237 weeks from 2006 to 2011.  

\section{Results}
All uncertainties given are the standard error of the mean (SEM).\\

\subsection{Receiving Points}
\label{sec:rescaling}
\begin{figure}[!ht]
    \centering
    \includegraphics[width=.75\linewidth]{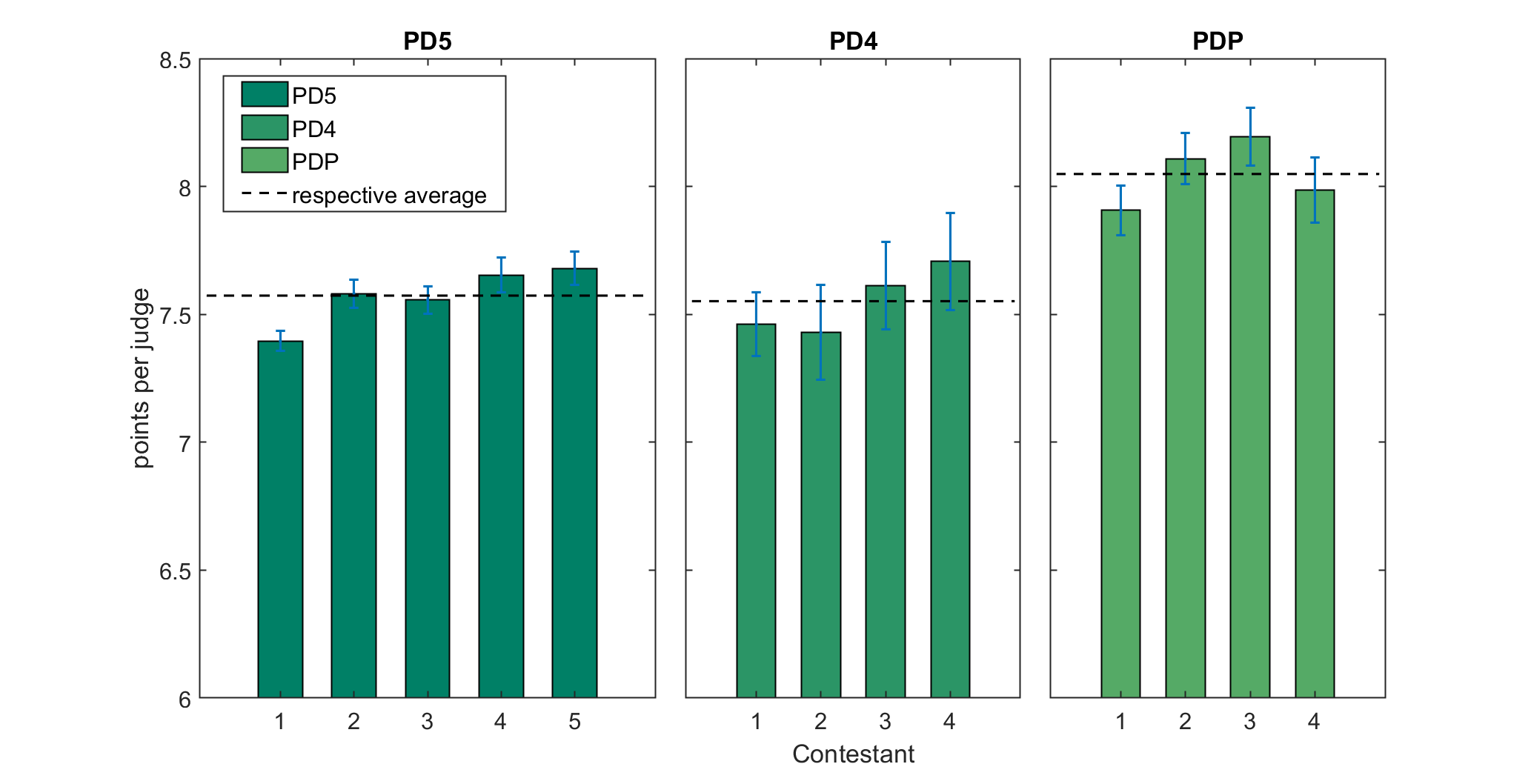}
    \caption{The average amount of points received per judge for each contestant for the regular show PD5 (left), with only four contestants PD4 (middle) and the celebrity version PDP (right).}
    \label{fig:points_orig}
\end{figure}

The average points received per judge for each day for the different data sets is shown in \autoref{fig:points_orig}.\\
\vspace{0.5cm}

\begin{figure}[!ht]
    \centering
    \includegraphics[width=.75\linewidth]{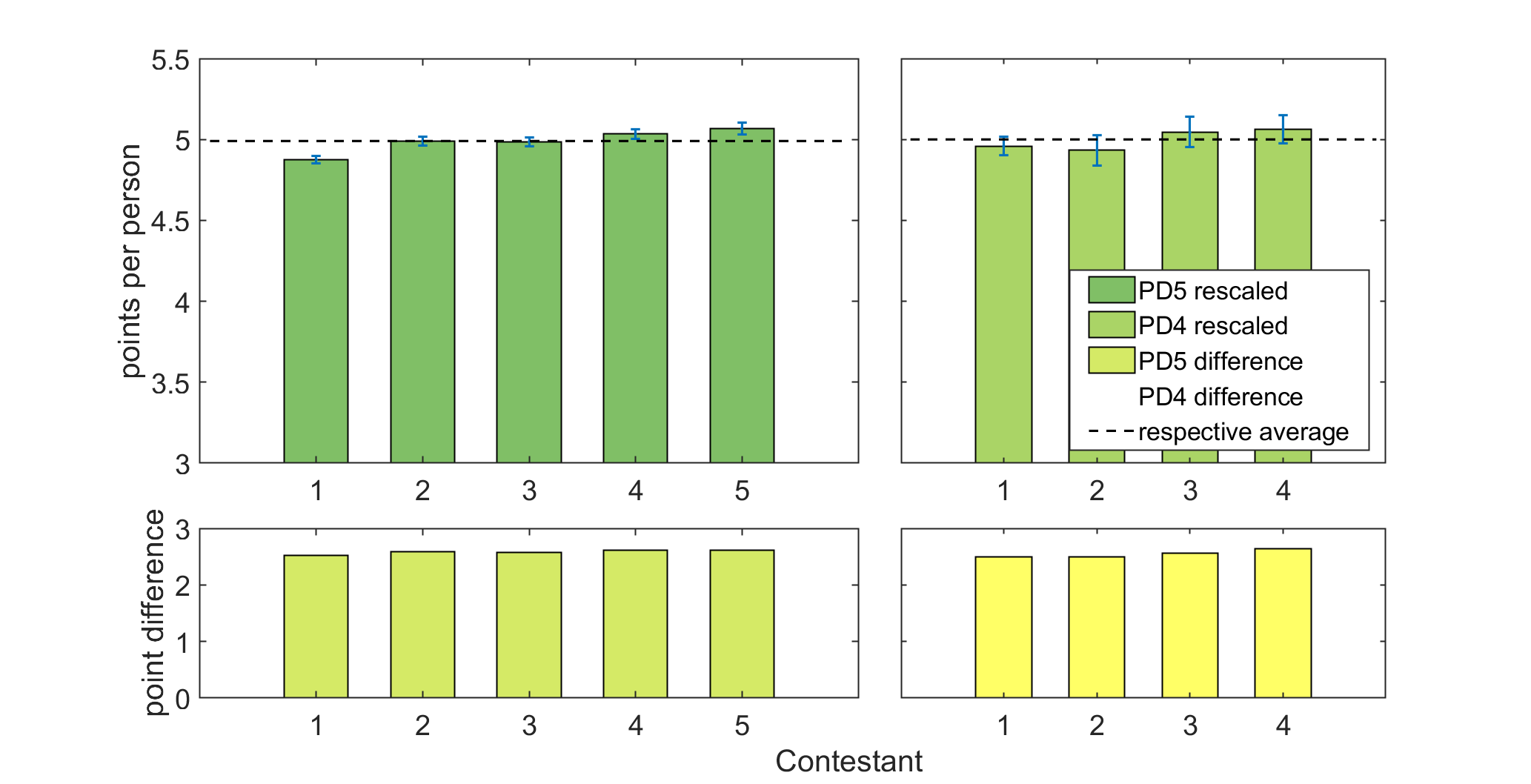}
    \caption{The average amount of rescaled points received per judge for each contestant for the regular show PD5 (upper left) and with only four contestants PD4 (upper right). The lower plots show the difference between the original and the rescaled data.}
    \label{fig:points_rec}
\end{figure}

Two observations can be concluded: (a) The average points received for the two regular shows are within uncertainty identical while the celebrity version will have on average a higher number of points given by 0.5. (b) In any data set, the first contestant receives on average less or the least amount of points. For PD4, the second contestant receives even less points on average then the first, but with a larger spread.\\
As motivated and discussed in \autoref{sec:method}, we rescaled the data to try to factor out subjective grading. This was only possible for the data set of PD5 and PD4 where the individual votes were available. The result of this approach is shown in \autoref{fig:points_rec}.\\ 
\vspace{0.5cm}

As expected, the overall structure of the distribution was not changed since only average values are shown. 
Each contestant loses about the same amount of points. It has to be noted that the third contestant in PD5 and the second contestant in PD4 lose slightly less than average points, which will have a great impact on the winning distribution.

\vspace{0.5cm}

\subsection{Winning Probabilities}

\begin{figure}
    \centering
    \includegraphics[width=0.8\linewidth]{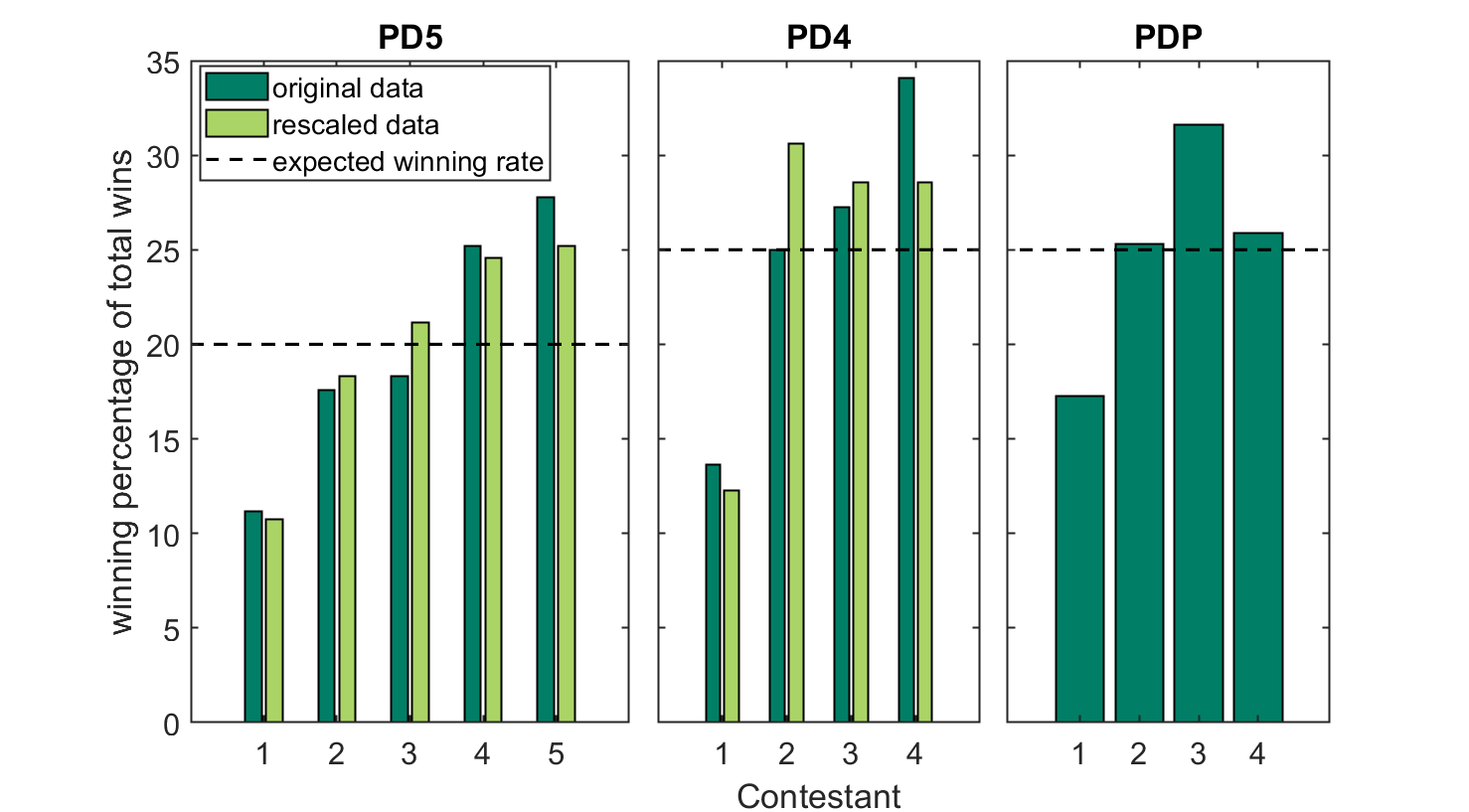}
    \caption{Winning probability for each day according to the original and the rescaled data for the three data sets.}
    \label{fig:winners}
\end{figure}
\vspace{0.5cm}

Given the average points received, we expected the first candidates to be the least likely to win and the others to have roughly the same odds, as you can see in \autoref{fig:winners}. In addition, the general rule 'the latter a candidate performs, the more likely he is to win' was found to be true. It has to be noted, that in PD5 this dependence is greater than in PD4. However, in the celebrity version it is not the last but the next to last candidate who is most likely to win, indicating a different voting behaviour.\\

\subsection{Multiple Winners}

\begin{figure}
    \centering
    \includegraphics[width=.8\linewidth]{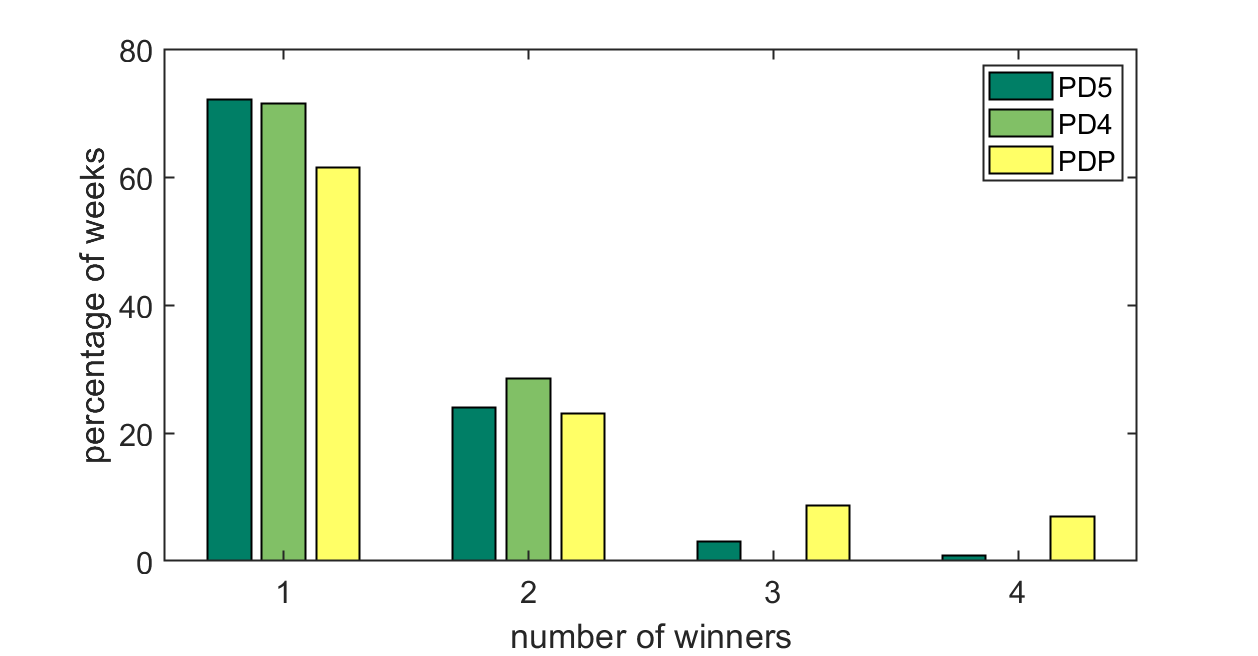}
    \caption{Percentage of n winners in a single show.}
    \label{fig:winnerdistribution}
\end{figure}

Due to the discrete voting system, multiple winners per week are possible. \autoref{fig:winnerdistribution} shows the percentage of how often n winners in a single contest have been found.

\vspace{0.5cm}

The greatest number of multiple winners is found in PDP, especially 3 and 4 winners are far more common than in the regular show. Observing this fact, we also had a look at the different pairings of winners in PDP, as you can see in \autoref{tab:wins}.  

\begin{table}[hbt]
\centering
\caption{statistics of the wins at Promi-Dinner (PDP)}
\vspace{0.5cm}
\begin{tabular}{ll|cccc}
\toprule
\multicolumn{2}{l|}{winning}                         & \textbf{1} & \textbf{2} & \textbf{3} & \textbf{4} \\ 
\cline{1-6}
\multicolumn{2}{l|}{\textbf{alone}}                  & 15         & 26         & 35         & 31         \\
\cline{3-6}
\multirow{4}{*}{\textbf{together with}} & \textbf{1} & -          & 8          & 11         & 5          \\
                                        & \textbf{2} & 8          & -          & 12         & 9          \\
                                        & \textbf{3} & 11         & 12         & -          & 8          \\
                                        & \textbf{4} & 5          & 9          & 8          & -          \\
 \bottomrule
\end{tabular}
   \label{tab:wins}
\end{table}
For PDP, the first contestant wins alone in just 38\% of the cases, the others significantly more often, about half of the time. The last contestant wins mostly alone, which coincides with the least total amount of wins. Furthermore, there are 8 occurrences of three or four winners on PDP, these winners averaged 8.8 $\pm$ 0.3 points, so below the overall winning average. \\ 
In addition,comparing contestant 3 to the others, he wins more often alone than contestant 2 and more often together than contestant 4, resulting in the most wins overall (see \autoref{fig:winners}).

\subsection{Winner-Loser Difference}
\begin{figure}
    \centering
    \includegraphics[width=\linewidth]{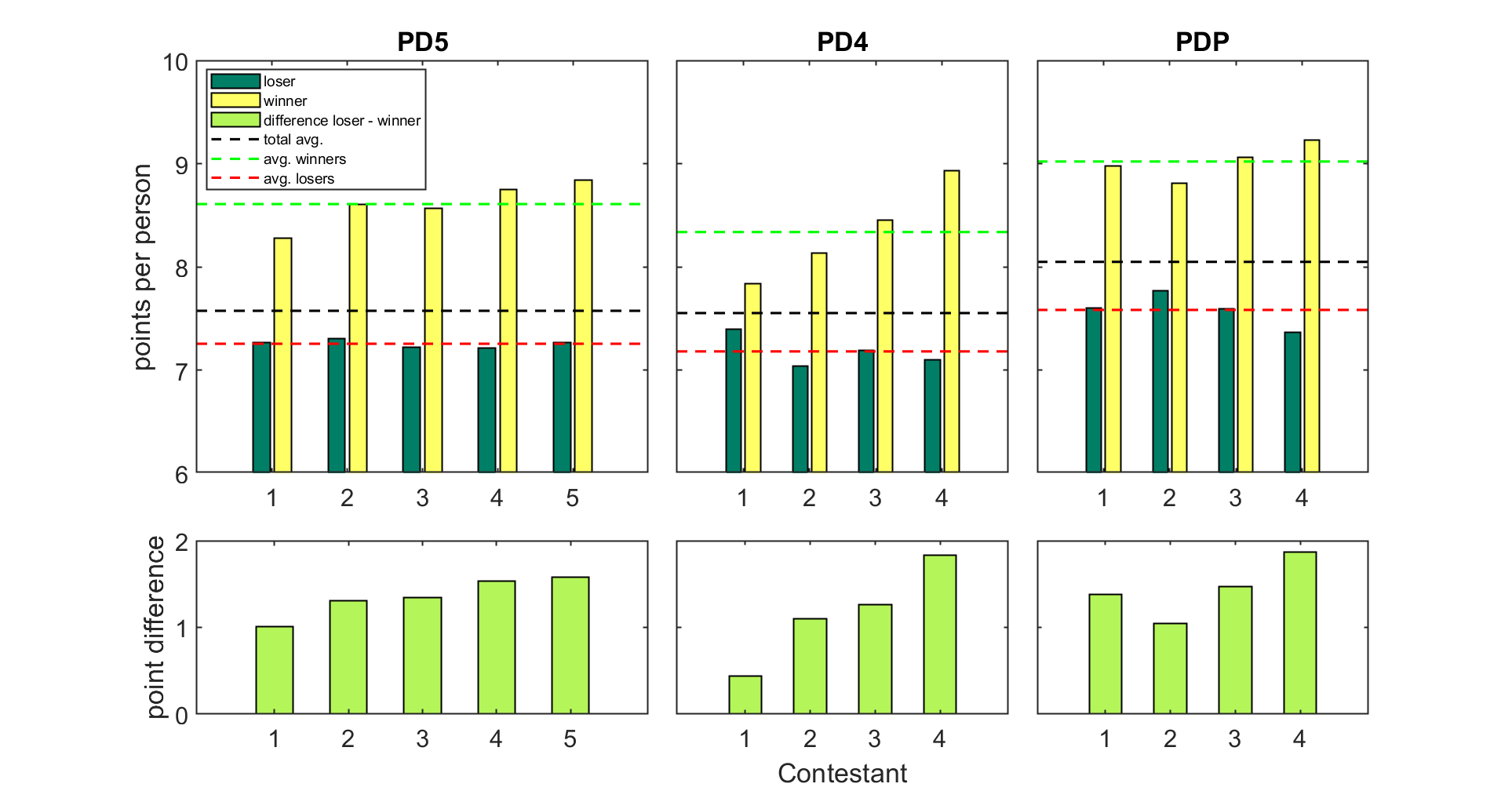}
    \caption{Average points of a contestant in case he wins and average not-winning points of the other contestants which then lost in the same show.}
    \label{fig:distr_orig}
\end{figure}

\begin{figure}
    \centering
    \includegraphics[width=\linewidth]{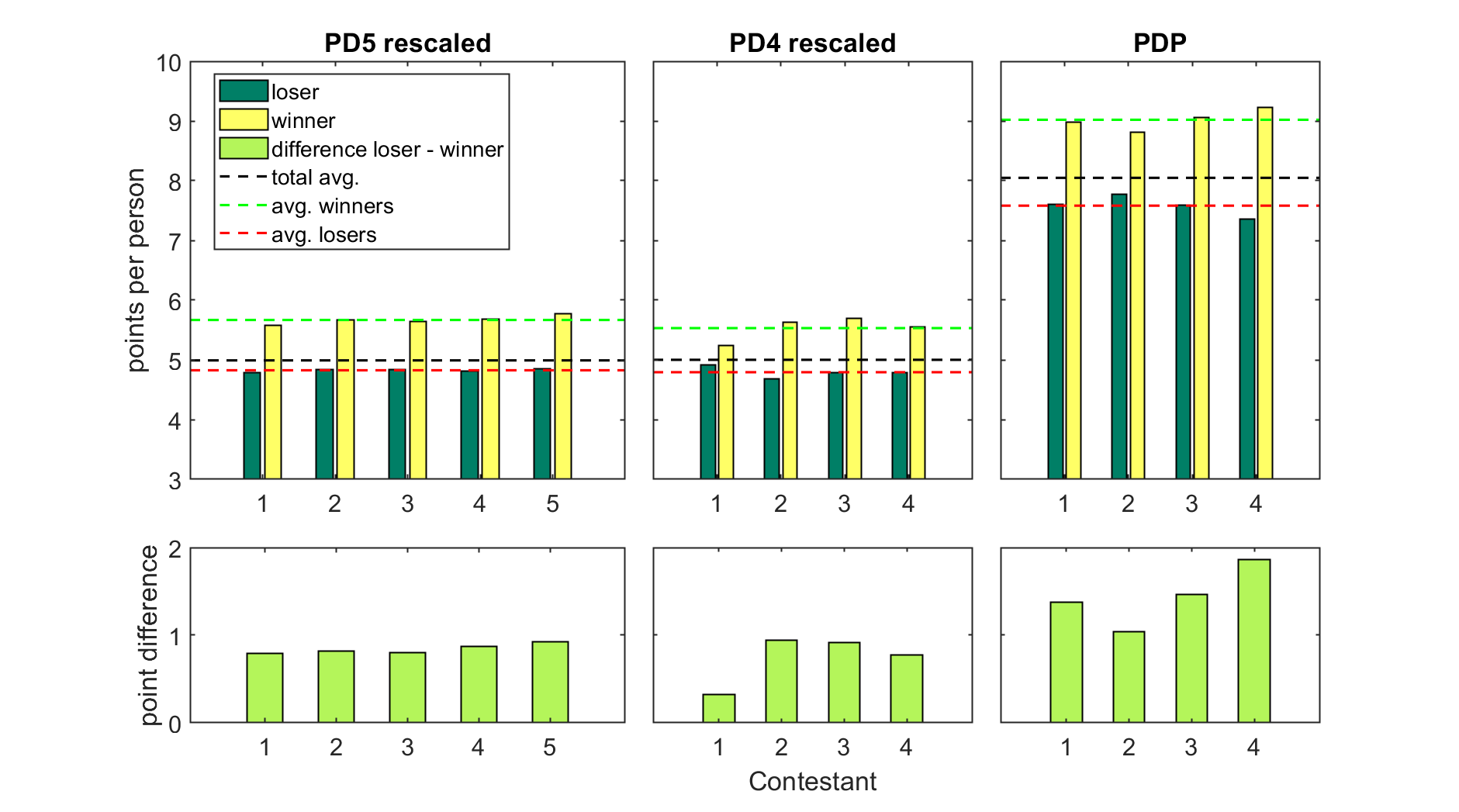}
    \caption{Points of the winner and the losers of the rescaled data broken down by contestant.}
    \label{fig:distr_rescaled}
\end{figure}

To further understand the winning probabilities, we looked at the distribution of the average winning points compared to the then losing points. Therefore we calculated the average points received in the case of winning and the average points of the other losing contestants as you can see in \autoref{fig:distr_orig} and \autoref{fig:distr_rescaled}.\\
\vspace{0.5cm}

\vspace{0.5cm}
For the two regular shows PD5 and PD4, the order effect is observed again in the winner points distribution, while the respective losing points is nearly constant for PD5 and only a slight excess is observed for Mondays in PD4. In both shows, in the case when the first contestant wins, the advance is smaller than for the other winners. 
The points in PDP shows a different trend - the first contestant tends to win with an exceptionally great lead. This observation is in contradiction to the first contestant being the least probable to win and therefore clearly shows a different voting behaviour between the normal and celebrity shows.\\

\subsection{Awarding points}
\begin{figure}
    \centering
    \includegraphics[width=\linewidth]{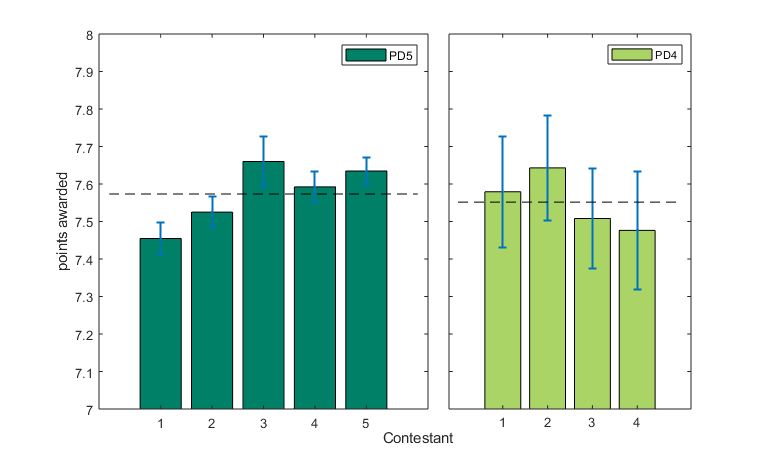}
    \caption{Average points awarded by each contestant}
    \label{fig:award}
\end{figure}

We further had a look at the points given by each contestant (\autoref{fig:award}). The average value matched of course the average received value. For PD5 we observed again an order effect, except that the third contestant tends to give more points than any other. The results for PD4 are not conclusive, since the SEM is to large.

\begin{figure}
    \centering
    \includegraphics[width=.6\linewidth]{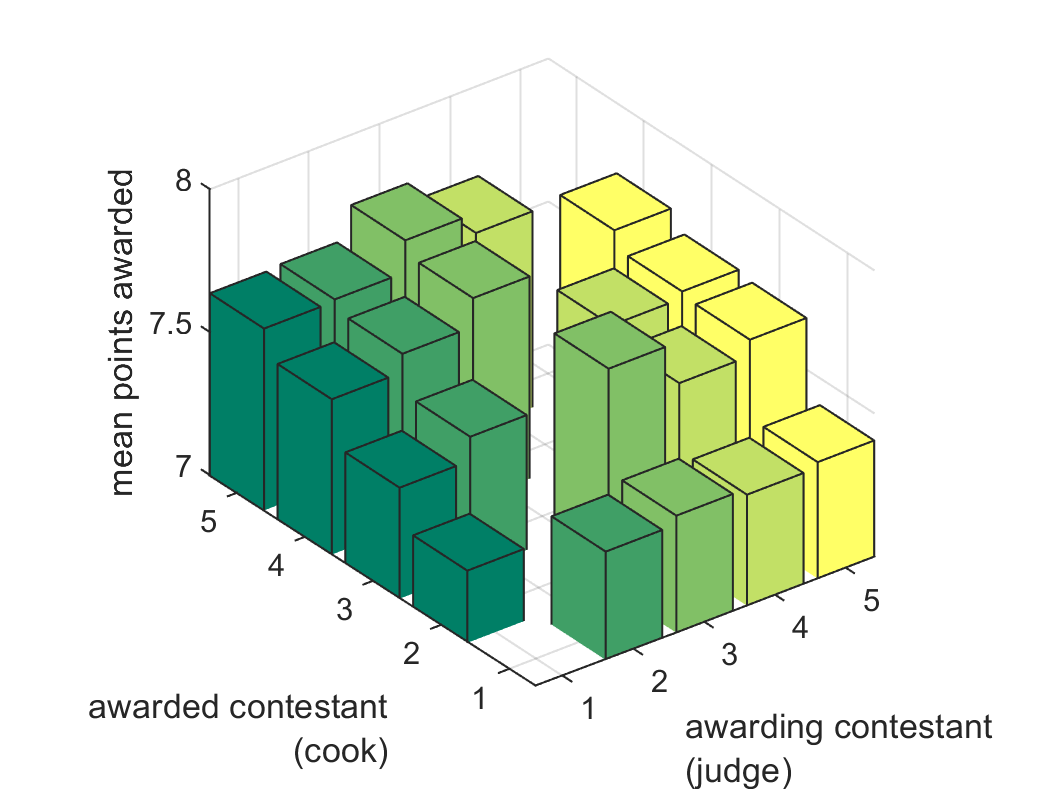}
    \caption{Average points given by contestant X to contestant Y}
    \label{fig:award_indiv}
\end{figure}

To further investigate the 'generosity' of the third contestant, we had a closer look at PD5 and calculated and overall voting matrix for the person to person ratings, e.g. how many points contestant 1 awards to contestant 2 , 3  and so on. The result is shown in \autoref{fig:award_indiv}.\\
\vspace{0.5cm}

The results shown in \autoref{fig:points_orig} and \autoref{fig:award} are the projections onto one axis. Again, the later a contestant performs, the more points he receives. This holds also true for the individual ratings by each judge. Also, the later a contestant performs, the more points he awards. The only exception from these trends is contestant 3 awarding significantly more points to 2 than expected. This is the underlying effect for two observations: the second contestant receiving more than the order effect would allow to expect seen in \autoref{fig:points_orig} and the excess of points awarded by the third contestant giving in \autoref{fig:award}.\\

It should be mentioned, that the first contestant is receiving almost equally low votes by everyone and that the lowest average amount of points is received by the second contestant from the first.

The data of PD4 is inconclusive and due to the large uncertainty hard to interpret, but it can also be seen that the third contestant awards more than expected points to the second contestant.

\subsection{Cooking bias}
\label{section:cooked_bias}

It was proposed in \textsuperscript{\cite{schueller}} that the cooking order also affects the points in such a way that if a person has already cooked, it is suggested that the person becomes more critical with the others and awards less points on average. In the data of the regular show, we could not verify this statement and even observe the opposite: A positive influence on the points awarded. The applied method is, that we split the awarded points by a cook in two sets: whether the person had already cooked or not. The average points awarded as well as the ratio between the two averages can be seen in the tables \ref{tab:cookedPD5} and \ref{tab:cookedPD4}.\\

\begin{table}[htb]
\centering
\caption{Averages of the given points prior and after cooking for PD5 and the ratio between the two average values.}
\vspace{0.5cm}
\begin{tabular}{crrr}
\toprule
\multicolumn{1}{c}{Contestant} & \multicolumn{1}{c}{prior to cooking} & \multicolumn{1}{c}{after cooking} & \multicolumn{1}{c}{ratio a/p} \\
\cline{1-4}
1 & -                & 7.47 ($\pm$0.04) & -                \\
2 & 7.38 ($\pm$0.05) & 7.60 ($\pm$0.05) & 1.03 ($\pm$0.01) \\
3 & 7.59 ($\pm$0.08) & 7.77 ($\pm$0.10) & 1.02 ($\pm$0.02) \\
4 & 7.57 ($\pm$0.04) & 7.76 ($\pm$0.07) & 1.03 ($\pm$0.01) \\
5 & 7.63 ($\pm$0.04) & -                & -                \\ 
\bottomrule
\end{tabular}
\label{tab:cookedPD5}
\end{table}

\begin{table}[htb]
\centering
\caption{Averages of the given points prior and after cooking for PD5 and the ratio between the two average values.}
\vspace{0.5cm}
\begin{tabular}{crrr}
\toprule
\multicolumn{1}{c}{Contestant} & \multicolumn{1}{c}{prior to cooking} & \multicolumn{1}{c}{after cooking} & \multicolumn{1}{c}{ratio a/p} \\
\cline{1-4}
1 & -                & 7.58 ($\pm$0.15) & -                \\
2 & 7.60 ($\pm$0.16) & 7.67 ($\pm$0.17) & 1.01 ($\pm$0.03) \\  
3 & 7.50 ($\pm$0.14) & 7.52 ($\pm$0.23) & 1.00 ($\pm$0.04) \\
4 & 7.48 ($\pm$0.16) & -                & -                \\
\bottomrule
\end{tabular}
\label{tab:cookedPD4}
\end{table}
While both tables show that the average points awarded after cooking are almost always higher than the ones awarded prior to cooking, only the PD5 results are significant. PD4 shows the same trend but the data set is to small to be significant. Hence, the results support the interpretation of cooking having a positive bias on the voting behaviour.  

\newpage
\section{Discussion}

Our first hypothesis,which is stating that an 'order effect' exist, can on one hand be seen in the first candidate winning the least, which was already stated by \cite{haigner}, but on the other hand also by the clear dependency of the winning rate on the order of candidates. This effect is caused by several reasons.\\

Firstly, the amount of points received depends on the cooking position. The data implies a proportionality: the later a contestant performs, the more points he receives (c.f. \autoref{fig:points_orig}). One explanation for this voting behavior could be that the contestants get to know each other better over the course of the week, hence awarding each other more points out of sympathy. As the first contestant is receiving the least amount of points, he is less likely to win.\\
It has to be noted that there are some disagreements with the strict proportionality, especially the last contestant in PDP receiving less than average points, these will be further discussed in \autoref{section:PDP}. In addition, this sympathy voting is not supported by the PDP data.\\

Secondly, similar to the optimal solution strategy of the secretary problem, the first contestant will act as a reference mark. No matter his performance, he will receive a thought to be average amount of points to keep an option for future contestants. This behavior can be seen in \autoref{fig:award_indiv}, as the first contestant is graded equally low by the others. This is an adapted version of the secretary strategy. A complete rejection is not necessary because of the rating system: The contestants rate an in their thought average amount of points to the first contestant. As you can see in the data, the amount of points turns out to be less than average, as the later contestants will mostly receive higher votes as they can be compared. The first contestant is used as a reference mark which will most likely be not undercut. In the end, people will give more points because of the presented reasons and also to emphasize their vote for a good contestant. This can be seen not only in a higher number of points received, but the later contestants also win with a greater lead (c.f. \autoref{fig:distr_orig}).\\

Moreover, contestants award more points after they have cooked. In PD5 our data show an improvement of approx. 0.2 points after cooking. This result may be a reason of underestimation of the own performance, but also due to forgiving little mistakes of the others bearing in mind one's own performance. Another possible explanation could be an over-confidence of the contestant right after cooking (\cite{schueller}).  Furthermore, the calculated averages are shifted by the other observed phenomenons as the points awarded after one's cooking are typically given later in the week.\\
The observed correlation is contrary to  \textsuperscript{\cite{schueller}}, who observed a point reduction after cooking. Supporting that point of view, the data shows that especially the first and second contestant award less than expected points directly after cooking (\autoref{fig:award_indiv}). So while the proposed 'cooking bonus' was found for the averages, it might not hold true for the first contestant after one's cooking. This disagreement might be explained by the simple amount of available data. In \textsuperscript{\cite{schueller}} only 237 weeks from 2006 to 2011 were used, while in our analysis 597 weeks from 2006 to 2019 were used. If this difference between the observations is due to an evolution of the voting behaviour of the years or simply a matter of statistics is not clear.
With this said, we can not finally decide whether there is a 'cooking bonus' or not. Our calculated averages are subject to the other described phenomenons but \textsuperscript{\cite{schueller}} trying to avoid these using the method of multi variable linear regression, may have produced a non-existing correlation, as the proposed negative bias is always counter-acted by other tendencies.\\

\subsection{Secretary Problem}
To determine the best cook from four or five candidates, the optimal strategy based on the secretary problem suggests rejecting the first candidate and choosing the next one better than him. Although the average contestant supposedly does not know this strategy, the data suggests that the contestants already apply a similar strategy. The winner distribution (\autoref{fig:winnerdistribution}) clearly shows the first contestant to be rejected most often, which would also be the case applying the secretary strategy.\\

If one would apply the secretary strategy rigorously, the last contestant would be twice as likely to win. This is not seen in the data. One reason for this might be that the rating system is continuous. One does not have to choose the last candidate when one has rejected all other candidates before, especially since rejecting in this sense is not existing. Still, some kind of rejecting (e.g. lower votes) until someone is better results in a greater emphasis on a good contestant later in the game to emphasize the judges decision. The contestants are rated equally low, until someone is better, but if someone even better comes around, he will get even more points. This model would also result in an increasing winning rate over contestants and an increasing winning lead. Both was observed in the data\\

Hence, the secretary problem, if adapted to a continuous rating system, seems to fairly well coincide with the data. This implies a subconscious understanding of a good method of finding one of the best candidates.

\subsection{Rescaling of the Data}
The rescaling was done to reduce the influence of subjective ratings and therefore the over-proportional impact of the "outlying" voters (c. f. \autoref{sec:rescaling}). The shape of the distribution of the awarded points (\autoref{fig:points_rec}) does not change much after rescaling. However the winner distribution (\autoref{fig:winnerdistribution}) is different for the rescaled data. Additionally, the two rescaled data sets have in common that the first contestant and last contestants win less often compared to the original data, whereas the others improve. Furthermore, the increase in the winner's lead over time is not observed in the rescaled data. \\

The different winning distribution suggests that the subjective rating is in favor of the later contestants. This coincides with the increasing winning point difference, suggesting more generous people give more points to later performing contestants. Therewith the later contestants get less points during the rescaling, resulting in a lower winning percentage.\\
Especially the double winning cases are crucial for the later winners, as the rescaling almost completely eliminates multiple wins.\\

The "winners lead increasing over time" seems also to depend on the subjective ratings, as it is not visible when the ratings are removed. This observation against expectations raises the question whether this is all due to the subjective ratings being removed or whether there are other mayor influences introduced through the rescaling.\\

During the rescaling we assumed the awarded points to be equally distributed amongst the judges, which was later found out to be not the case (\autoref{fig:award}). Especially contestant 3 in PD5 and 2 in PD4 award significantly above average. Therefore these highly awarding contestants are more affected by the rescaling in terms of point reduction. Hence the others get less points and the highly awarding contestant an advantage, which can result in his favour in the case of close or equal points. This analogously holds a disadvantage for the lowly awarding contestants.\\

To conclude, the used method of rescaling removes the subjective grading but also introduces an effect because of the awarding distribution. Both effects change the results in a similar order of magnitude. So it can not be said which effects occur because of which source.\\
The issue could be solved by using a different rescaling method considering the awarding distribution. However, to do this, one would need a bigger data set to minimize the error on the distribution.\\

\subsection{Voting in PDP}
\label{section:PDP}
The statistics of the Promi-Dinner show a slightly different voting behavior compared to the regular show.
\begin{itemize}
    \item a higher average of the points awarded
    \item the last contestant receives less points than expected and wins less likely.
    \item many multiple winners per week
    \item The first contestant wins with a significantly larger lead than in PD4 and PD5
\end{itemize}
The premise of the celebrity version is a different, as the celebrities do not play for themselves but for charity. In addition, the celebrities may know each other and play a public role which they try to influence. Therefore their merit is not the money rather than the fame and coverage. So they do not need to win but make a good performance. They have no reason to give bad ratings, which then here results in a higher average points received. Pushing the point spectrum more to the upper end, the differences in points become smaller and thus multiple wins more likely (\autoref{fig:winnerdistribution}). \\
Additionally, the celebrities mostly know each other before the show and therefore the bonus for the later contestants due to knowing each other better mentioned above applies directly from the start.\\

In PDP the third contestant has the greatest chances of winning as he wins more often alone than 2 and more often together than 4. This observation clearly shows the differences in voting behaviour between the normal and the celebrity version.

\subsection{Further Observations}
Comparing the two data sets of the regular show, it has to be noted that the first contestant is less likely to win in PD4 and if he does, it is with a exceptionally small lead. This could be due to the reduced number of contestants. Everyone has a greater chance of winning since 4 and not 5 contestants participate, and therefore especially the last contestants award less points to boost their chances. This disadvantage for the first contestant makes him to win less and closer.\\

Next to the first contestant winning the least,in PD5 the third contestant awards significantly more points to 2 , which can directly and indirectly be observed in several figures. A similar voting behaviour also has been observed in PD4, although non-significantly. A suitable explanation for this behavior has neither been found in the data nor in literature.\\

\section{Conclusion}

The show of 'Das perfekte Dinner' is not a fair game. The likeliness of winning strongly depends on the position of cooking, and not only for the first, but for all contestants. The comparison to the secretary problem showed, that although this optimal strategy is not commonly known, people subconsciously apply it in their voting behavior. So to come back t our original competition,  the success in competitions like poetry or science slams strongly depends on the position of performing, giving a great disadvantage to the first performer and an advantage to the last. So when competing one should always carefully choose when to compete.

\end{document}